\documentclass[conference]{IEEEtran}
\usepackage[utf8]{inputenc}
\usepackage{amstext}
\usepackage{graphicx}
\usepackage{siunitx}
\usepackage{float}
\usepackage[tableposition=top]{caption}

\floatstyle{plaintop}
\restylefloat{table}
\usepackage{url}
\usepackage{xcolor}

\begin{document}

\title{Impact of Mobility-on-Demand on Traffic Congestion: Simulation-based
Study}

\author{
\IEEEauthorblockN{David Fiedler, Michal \v{C}\'{a}p and Michal \v{C}ertick\'{y}}
\IEEEauthorblockA{
Department of Computer Science,
Faculty of Electrical Engineering,
CTU in Prague, 
Czech Republic
}
}

\maketitle

\begin{abstract}
The increasing use of private vehicles for transportation in cities results in a growing demand for parking space and road network capacity. 
In many densely populated urban areas, however, the capacity of existing infrastructure is insufficient and extremely difficult to expand. 
Mobility-on-demand systems have been proposed as a remedy to the problem of limited parking space because they are able to satisfy the existing transportation demand with fewer shared vehicles and consequently require less parking space. 
Yet, the impact of large-scale vehicle sharing on traffic patterns is not well understood. 
In this work, we perform a simulation-based analysis of consequences of a hypothetical deployment of a large-scale station-based mobility-on-demand system in Prague and measure the traffic intensity generated by the system and its effects on the formation of congestion.
We find that such a mobility-on-demand system would lead to significantly increased total driven distance and it would also increase levels of congestion due to extra trips without passengers. In fact,  38\%\,kilometers traveled in such an MoD system would be driven empty.  
\end{abstract}

\section{Introduction}
In many densely-populated urban areas, the use of privately-owned automobiles as the main mode of transportation is unsustainable. Private vehicles take up large amounts of space when parked at a destination, require high-capacity highways, and significantly contribute to air pollution. 
Mobility-on-Demand (MoD) systems can help reduce the parking space requirements by increasing the utilization of vehicles in the system.
A MoD system consists of a fleet of vehicles shared by users of the system to realize their mobility needs. 
Taxi service, Uber, or Lyft are examples of conventional MoD systems; the affordability of the service can be further improved by employing the self-driving technology~\cite{avip2017}. 
Since a typical shared vehicle serves more passengers during a day than a typical personal vehicle, the same travel demand can be satisfied with fewer shared vehicles than with personal vehicles. The work presented here suggests that a properly designed MoD system can achieve up to fivefold reduction in the number of required vehicles. Moreover, due to higher utilization, the cars in the system spend
less time parked. 
Since MoD systems consist of fewer vehicles, each spending less time parked, mass adoption of this transportation paradigm has a potential to drastically reduce demand for urban parking space. 

Despite having the clear benefit of reducing the parking space requirements, it is also important to understand the impact of large-scale vehicle sharing on total traffic intensity and on formation of traffic congestion.
Indeed, in order to transport all the passengers, shared vehicles need to cover at least the same distance as conventional personal vehicles. 
However, unlike personal vehicles, shared on-demand vehicles must also travel empty between the individual passengers, i.e., from the destination of the current passenger to the origin of the following passenger. 
On top of that, passenger demands are not distributed uniformly in time and space, which causes vehicles to accumulate in some parts of the city and while other areas may suffer from the shortage of vehicles.
To ensure that the each region of a city has a sufficient number of available vehicles, the \emph{rebalancing} has to be implemented, i.e. vehicles must travel empty from areas with a surplus of vehicles to areas with a shortage of vehicles~\cite{Spieser_2014-AMoD-Singapore}.
Therefore, on-demand transportation increases the total amount of traffic in the system. 
In this paper, we use micro-simulation to investigate the extent of this effect by measuring the impact of empty trips on total vehicular traffic and formation of traffic congestion. 

\subsection{State of the art}\label{subsec:sota}
The properties and performance characteristics of MoD systems have been previously studied either using simulation or by formal analysis of mathematical models of such systems.   

In~\cite{Fagnant2014}, for example, a multi-agent simulation was used to measure the impact of deployment of a small fleet of MoD vehicles, and the results suggest that the fleet can be reduced nearly twelve times, but the total distance traveled increases by 10.7\%. The authors also analyze the environmental impacts such as the number of cold starts, founding that the emission savings would be significant. 
Another simulation study~\cite{Marczuk2016} compares the number of cars needed when using different vehicle rebalancing methods. In results for offline and online rebalancing, 28\% and 23\% fewer cars are needed compared to the system without rebalancing.

The congestion effects of rebalancing in MoD system have been numerically explored in \cite{zhang_control_2016} leading to the following structural insight: Although the need for rebalancing in MoD systems generates a significant amount of extra traffic, the rebalancing vehicles travel in the opposite direction than the vehicles carrying passengers, and thus they use lanes that are currently underutilized. Numerical experiments in simple synthetic networks indeed confirm the observation, suggesting that if the demand traffic flows (vehicles carrying passengers) are below road capacity limit, then the rebalancing flows (rebalancing vehicles) will also not exceed the capacity and consequently they will not introduce new congestion.
Then in~\cite{Zhang2016} the problem of vehicle routing in congested networks was modeled to formally study the impact of rebalancing on formation on congestion. Assuming symmetric network and time-invariant demand, the authors proved that if the demand flows can be routed without exceeding free-flow capacity, then the rebalancing vehicles can be also routed through the network without exceeding road capacities. Further, they performed experiments demonstrating that the method remains effective even in real-world networks that are not perfectly capacity symmetric.   
Also, the authors analyzed the system in a steady state, i.e., they work with the assumption that the travel demand is time-invariant. Clearly, real-world travel demand changes over time (e.g., compare travel demand in the morning peak and the demand at night), but, as the authors point out, if the demand intensities change slowly relative to the average trip duration, the model remains reasonably valid. In large urban areas, however,  the traffic intensity may rise drastically over the time period of a single trip. Consider, for example, a  30-minute car trip and the difference of traffic intensities  between 6:00 and 6:30, i.e., during the onset of morning peak.

Another aspect that has recently received significant attention is car sharing. For example, in~\cite{Pnas2016}~, authors study the impact of large-scale ride-sharing and demonstrate that ride-sharing can further reduce the number of vehicles needed to satisfy the travel demand and moreover drastically reduce the total traveled distance by the shared vehicle.

\subsection{Our Contribution}
In this paper, we study the impact of large-scale MoD system deployment on total traffic intensity and formation of congestion using simulation. 
Unlike other existing works, 1) we simulate the system over the period of entire day (24h), which is important to properly account for the changes in travel demand intensities, 2) we use real road network of Prague (cz), which is not capacity symmetric, 3) we use traffic demand data that is representative of real traffic demand in Prague both in structure and scale and 4) we simulate the entire lifetime of an on-demand vehicle including pick-up and drop-off trips. 

This higher-fidelity model allows obtaining new insights into the impacts of massive deployment of MoD systems. In particular: 1) the model highlights the contribution of pick-up and drop-off trips to the formation of congestion in station-based MoD systems and 2) it suggests that if the demand cannot be routed through the road network without exceeding the free-flow capacity, then the deployment of MoD system may actually worsen the congestion situation.      

\section{Background Material}
In traffic engineering, the effects of traffic congestion on a road segment are modeled by the \emph{fundamental diagram of road traffic}~\cite{Haight1963}, which relates traffic density to traffic flow.  \emph{Traffic density} is the number of vehicles per unit of distance of the road segment, while \emph{traffic flow} is the number of vehicles passing a reference point per unit of time. When traffic density grows, traffic flow also increases until it reaches a tipping point after which the flow starts dropping due to congestion. The tipping point is known as the \emph{critical density}~\cite{Kerner2009}. The exact shape of the diagram is typically determined by fitting empirical data from real-world observations of vehicular traffic. Subsequently, we will use the traffic density as a measure of utilization of a particular road segment and the critical density value  $y_{c}= 0.08\ \textit{vehicle}\,m^{-1}$  from~\cite{Tadaki2015}. The road segments with traffic density above critical
density $y_c$ will be referred to as \emph{congested road segments}.

\section{Methodology}
The effect of a hypothetical large-scale deployment of MoD in Prague is analyzed using simulation as follows. 
Firstly, using a demand model, we synthesize a set of trips that statistically correspond to trips by private cars performed during a typical work day. 
Then, we design a MoD system that is capable of serving those trips with a satisfactory quality of service. 
After that, we perform a simulation of two scenarios: 
1) In \emph{conventional system scenario}, all trips are served by private vehicles. 
2) In \emph{MoD system scenario}, the trips are served by a large MoD system. 
During simulation, we record information about road utilization and the service quality within the system. 
Finally, we compare the data collected in the two simulated scenarios.
We will now discuss each step in detail.

\subsection{Input data\label{subsec:Input-data}}
The set of trips that represent the transportation demand is generated by the multi-agent activity-based model of Prague and Central Bohemian Region introduced in \cite{certicky_fully_2015}.
In contrast to traditional four-step demand models~\cite{handbook-transport-modelling}, which use trips as the fundamental modeling unit, activity-based models employ so-called activities (e.g. work, shop, sleep) and their sequences to represent the transport-related behavior of the population. 
Travel demand is then occurring due to the necessity of the agents to satisfy their needs through activities performed at different places at different times. These activities are arranged in time and space into sequential (daily) schedules. Trip origins, destinations and times are endogenous outcomes of activity scheduling. 
The activity-based approach considers individual trips in context and therefore allows representing realistic trip chains. 

The model used in this work covers a typical work day in Prague and Central Bohemian Region. The population of over 1.3 million is modeled by the same number of autonomous, self-interested agents, whose behavior is influenced by demographic attributes, current needs, context, and cooperation. 
Agent's decisions are implemented using machine learning methods (e.g. neural networks and decision trees) and trained using multiple real-world data sets, including census data or travel diaries and similar transportation-related surveys. 

Planned activity schedules are subsequently simulated and tuned, and finally, their temporal, spatial and structural properties are validated against additional historic real-world data (origin-destination matrices and surveys) using six-step validation framework VALFRAM~\cite{drchal_data_2015}.
The model generates over 3 million trips per one 24-hour scenario. For our analysis, we selected the subset of 981\,775 trips that were performed in a private vehicle.

\subsection{On-demand system design}\label{subsec:mod_design}
\begin{figure}
\centering{}\includegraphics[width=1\columnwidth]{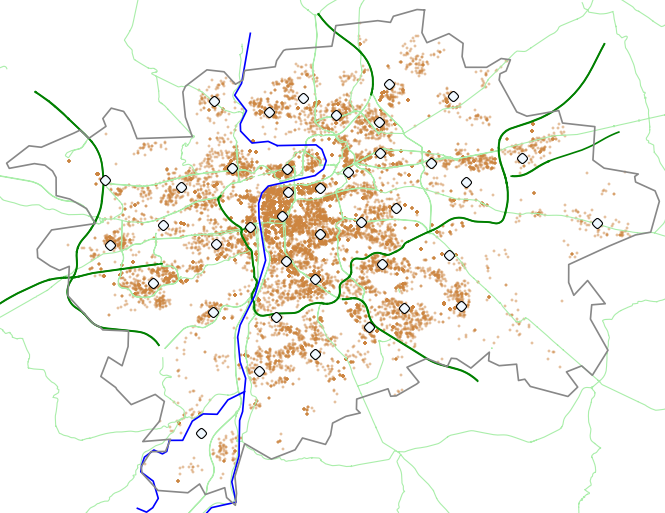}\caption{\label{fig:stations}MoD system stations in the city of Prague. Stations are shown as circles. Spatial distribution of origins and destinations of travel demands is depicted using brown dots.}
\end{figure}

In order to serve the trips generated by the demand model, we need to determine 1) the size of the on-demand fleet and 2) a policy for vehicle rebalancing within the system.
We adopt a station-based design of the MoD system~\cite{Spieser_2014-AMoD-Singapore}. That is, we partition the city into $n=40$ regions using k-means clustering over the demand data and we assume that there is a \textit{station} at the center of each such region. Stations serve as temporary parking lots for idle vehicles and they also contain facilities such as refueling/charging and cleaning. When a passenger requests a ride, a vehicle is dispatched from the nearest station to the passenger and drives to pick up the passenger. Then, it carries the passenger to its desired destination, where the passenger is dropped off. Finally, the vehicle parks again in the nearest station to the drop-off location. The result of this process is shown in Figure~\ref{fig:stations}. The number of regions was chosen such that the average travel time from station to a passenger is below 3 minutes.  

When a passenger requests a vehicle and the nearest station is empty, the customer cannot be served immediately and either has to wait for a vehicle to become available or a vehicle from a distant station must be dispatched, leading to long waiting time and consequently to customer dissatisfaction. Unfortunately, existing MoD systems are prone to such events due to the unbalanced structure of transportation demand. For instance, during the morning peak, the vehicles are typically requested for pickup in residential areas, but they subsequently end up in business districts. In results, the stock of vehicles in residential areas is shrinking, while unused vehicles are accumulating at business areas. This problem can be mitigated by continuous redistribution of vehicles from areas with a surplus of vehicles to areas with a shortage of vehicles.
This process is often referred to as rebalancing. In our hypothetical MoD system, we use linear programming based approach to balance the vehicle flows in the system~\cite{pavone_robotic_2012}. Due to time-varying demand and significant travel delays in the real-world system, the flows, however, cannot be balanced perfectly at every time point and consequently there must be a sufficient number of vehicles at each station to cover temporary shortages of vehicles. We determine the initial number of vehicles that are needed at each station experimentally, by deterministically simulating the evolution of the system and then by ensuring that each station has enough vehicles to never become empty. Using this process, we found that the system requires 51\,249 vehicles.  

\subsection{Multi-agent simulation}
In order to test the impact of the MoD system deployment on the road network utilization, we have implemented a multi-agent simulation of vehicular traffic in Prague based on  AgentPolis\footnote{https://github.com/agents4its/agentpolis} simulation framework. 

\begin{figure}
\centering{}\includegraphics[width=1\columnwidth]{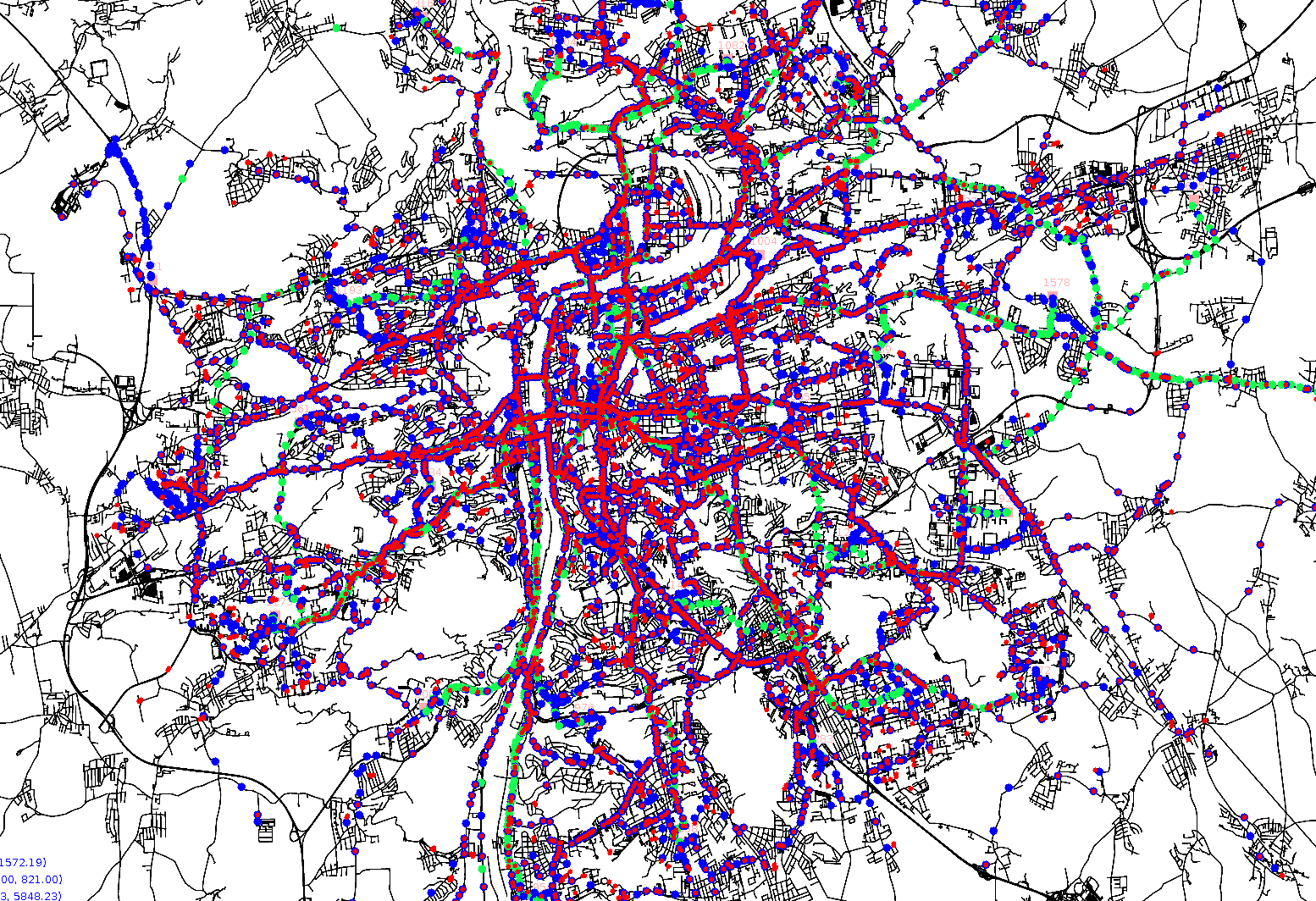}
\caption{\label{fig:ap}AgentPolis visualization of the traffic in Prague at 07:00.}
\end{figure}

AgentPolis is a large-scale multi-agent discrete-event simulation over a simulated environment that consists of: a) road network composed \emph{nodes} connected by road segments (\emph{edges}), b) stations in which on-demand vehicles park, c) on-demand vehicle agents, d) passengers agents.
The visualization of the whole city of Prague as simulated in AgentPolis is in Figure~\ref{fig:ap}. A video from the simulation is available at youtube.com\footnote{\url{https://youtu.be/e1SNCPrHo7w}}.

The topology of the road network is taken from OpenStreetMap\footnote{https://www.openstreetmap.org/}, resulting in a road network consisting of 306625 edges and 286009 nodes. 
The speed limit for all edges was set to \SI{40}{\km\per\hour}.

During initialization, we create the vehicles representing the on-demand fleet and assign them to appropriate stations (as computed during MoD system design phase). During simulation, at time points corresponding to travel demand from the input demand set, we create a passenger agent for the travel demand. The life cycle of one travel demand in the simulation is following:
\begin{enumerate}
\item A demand agent is created for the demand in the place where the demand trip starts. 
\item \label{itm:pickup} A vehicle from the nearest station is assigned to serve the demand. We plan the shortest path to the passenger and the vehicle drives it moves to the pick-up location. (pick-up trip). 
\item The vehicle picks up the passenger agent and carries the passenger to the desired destination location. (demand trip). 
\item The passenger agent is removed from the simulation.
\item The vehicle returns to the nearest station.  (drop-off trip). 
\end{enumerate}
If there is no available vehicle in the nearest station in point~\ref{itm:pickup}, the passenger agent is removed from the simulation, and it is considered as an \emph{unsatisfied demand}. 

Rebalancing schedule is computed at the on-demand phase and prescribes the number of empty vehicles to be moved between every two stations in each 10-minute time slice. Based on the rebalancing schedule, we order the required number of vehicles to drive between the two stations along the shortest path (rebalancing trip).

Every ten minutes, we count the number of vehicles of each category (pick-up, demand, drop-off, rebalancing) that drive on every road segment, which can be used to compute traffic density at those segments.

\section{Results}
\begin{figure*}
\centering{}\includegraphics[width=1\linewidth]{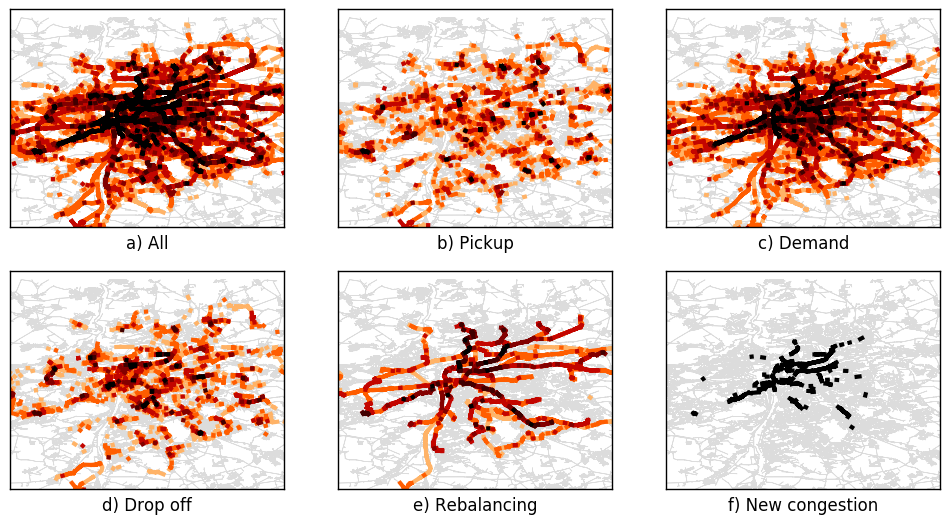}
\caption{\label{fig:main_map}Traffic density levels between 7:00 and 8:00. 
a) The traffic density after the deployment of the MoD system. 
b) Traffic density -- pick-up trips
c) Traffic density -- demand trips.
This image represents the traffic density levels on roads in the current state, without MoD system. 
d) Traffic density -- drop-off trips.
e) Traffic density -- rebalancing trips.
f) The roads on which the traffic exceeded the critical density after deployment of the MoD system.}
\end{figure*}

In this section, we discuss the result of our simulation. We will focus on the analysis of traffic patterns between 7:00 and 8:00, which is the time period with highest road network utilization. Figure~\ref{fig:main_map} shows heat maps depicting traffic densities on all roads in Prague for selected trip types. 
In particular,  Figure~\ref{fig:main_map}-a shows the overall traffic density in the MoD system.  Figure~\ref{fig:main_map}-b to \ref{fig:main_map}-e shows the traffic density contributions of different trip types. Note that Figure~\ref{fig:main_map}-c, in fact, shows the traffic density in the conventional system because the demand trips are the same trips that are driven by privately owned vehicles.  Comparing the Figure~\ref{fig:main_map}-a to \ref{fig:main_map}-c, we can clearly see that deployment of the MoD system would increase traffic intensity within the road network.  Figure~\ref{fig:main_map}-f shows the congestions introduced by the MoD system, i.e., the road segments that were below critical density in the conventional system (only counting contributions of demand trips), but exceeded critical density when we add the contribution of empty trips (pick-up, drop-off, and rebalancing). The number of lanes on each road segment was considered when computing the critical density, and the critical density was computed separately for each direction.

In Table~\ref{fig:traffic_km_share}, we list the share on the total distance traveled for each trip type. Note that the three MoD system specific trip types (rebalancing, pick-up, and drop-off trips) are responsible for more than a third of the total distance traveled (37.7\%).

\begin{table}
\scriptsize
\centering{}%
\begin{tabular}{|c|c|c|c|}
\hline 
Trip type & Avg. km/vehicle/day & Share on dist. & Share on cong.
\tabularnewline
\hline 
\hline 
Demand trip & 159 & 62.3\% & 61.3\%
\tabularnewline
\hline 
Pick-up trip & 31 & 12.1\% & 17.3\%
\tabularnewline
\hline 
Drop-off trip & 31 & 12.0\% & 14.4\%
\tabularnewline
\hline 
Rebalancing trip & 35 & 13.6\% & 6.9\%
\tabularnewline
\hline 
\end{tabular}
\caption{\label{fig:traffic_km_share}The share of the trip types on total travel distance and on congestion (share on traffic density on roads with traffic density higher than 50\% of the critical density).}
\end{table}

\begin{figure}
\centering{}\includegraphics[width=1\columnwidth]{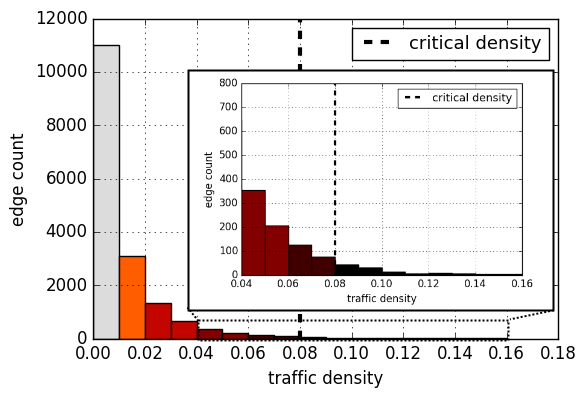}\caption{\label{fig:histogram_current_flow}Histogram of the traffic density in the current state (demand trips only). Edges with traffic density greater than 200\% of critical density ($0.16\,{veh}\,m^{-1}$) are represented by the last bin.
}
\end{figure}

Figure~\ref{fig:histogram_current_flow} shows the histogram of traffic density levels in the conventional system, before the MoD system deployment. For the simplicity of presentation, we excluded all road segments with traffic density lower than 1\% of the critical density ($0.0008\,\mathit{vehicle}\,m^{-1}$) from all histograms because they represent roads that are rarely ever used. 
These road segments represent 86.3\% and 82.3\% of all road segments for the conventional system and the MoD system respectively. To show a reasonable detail of the traffic density levels, there is also a focused histogram that contains only edges with traffic density greater than 50\,\% of the critical density~($0.04\,{vehicle}\,m^{-1}$).

\begin{figure}
\centering{}\includegraphics[width=1\columnwidth]{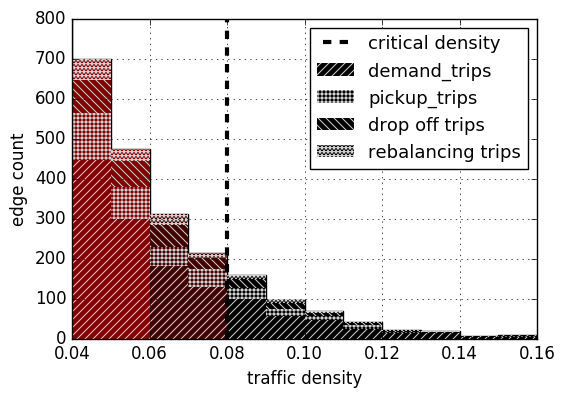}\caption{\label{fig:histogram_future_flow_detailed_stacked} Stacked histogram of the traffic density after MoD system deployment, created by the same methodology as the focused histogram in Figure~\ref{fig:histogram_current_flow}.}
\end{figure}

Figure~\ref{fig:histogram_future_flow_detailed_stacked} shows the histogram after deployment of the MoD system. It indeed confirms that there is a considerable increase of the traffic density. To investigate how different trip types contribute to the increased traffic density, we measured the share of traffic density per trip type for each edge and separated these shares into stacks. 
The data from this histogram was aggregated into Table~\ref{fig:traffic_km_share} for easier interpretation.

\subsection{Rebalancing}
We will now focus on rebalancing trips. The general insights into the congestion effects of rebalancing stipulating that rebalancing does not contribute to congestion because the rebalancing trips have the opposite direction than other traffic and thus the rebalancing vehicles use underutilized road lanes was partially confirmed. As we can see in our results, rebalancing has only 6.9\% share on congestion, despite having 13.6\% share on total distance driven (Table~\ref{fig:traffic_km_share}).

\subsection{Pick-up and drop-off trips}
Pick-up and drop-off trips have a significant effect on both travel distance and congestion. 
Our results show that these two trip types together are responsible for almost quarter (24.1\%) of the total distance traveled and more importantly, they have nearly a third of the share on total congestion (31.7\% -- Table~\ref{fig:traffic_km_share}). 
Although this number can be probably reduced by increasing the number of the MoD stations,
it is clear that the contribution of pick-up and drop-off trips cannot be ignored when designing an MoD system.

\section{Discussion}
The simulation results indicate that all types of empty trips contribute significantly to both total traveled distance and congestion. 
The impact on traveled distance is inevitable and can only be reduced by increasing the number of stations, which shortens pick-up and drop-off trips, or vehicles, which reduces the need for rebalancing.
The impact of the empty trips on congestion is more interesting. 
Pick-up and drop-off trips have a significantly larger share on traffic density than the share on traveled distance.
This effect can be explained by the fact that these trips are not distributed uniformly, on the contrary, they are accumulated creating clusters around the MoD system stations (see Figure~\ref{fig:main_map}-b and Figure~\ref{fig:main_map}-d ).

Although the contribution of rebalancing trips to congestion is disproportionately smaller than the contribution to total distance driven, it is still significant. 
This is perhaps surprising, since previous numerical and experimental results~\cite{zhang_control_2016,zhang_routing_2016} may suggest that rebalancing does not increase congestion. 
We could attribute this discrepancy to the invalidity of assumptions used by \cite{zhang_routing_2016} in our setting, but there are other possible explanations. 
First, our routing is not "congestion-aware", vehicles drive to their destination using the shortest path. This setup is different from the referenced work, where the routes of rebalancing vehicles are coordinated to prevent the formation of congestion. 
Second, the result from \cite{zhang_routing_2016} states that if the passenger flows can be routed through the network without exceeding road capacities, then vehicle rebalancing can be also routed without exceeding road capacities. As our results (and everyday experience) shows, the traffic network is unfortunately significantly congested already at the current state, i.e., under the conventional system. Therefore, even if the rebalancing flows were optimally routed, they may exceed the road free-flow capacities leading to increased congestion.
So far, we can only guess how much is the 6.9\% share of rebalancing on congestion caused by non-optimal routing and initial congestion, and how much it is caused by the fact that the simulated scenario does not satisfy the assumptions of the above theoretical analysis.

Another aspect that we did not model is the congestion effect in the simulation. In real traffic scenario, congestion tends to spread over the network and affect roads whose traffic density is not above the critical density. 
This effect is not implemented in our simulation, and thus our results underestimate the scale of congestion.
Moreover, we simulate the traffic flow as an uninterrupted flow, i.e. we do not model travel delays on intersections.

\section{Conclusion}
Large-scale deployment of mobility-on-demand systems in cities can dramatically reduce the number of vehicles needed to satisfy existing transportation demand and consequently reduce the need for parking. The impact of such systems on road utilization and traffic congestions is however relatively less understood. In this paper, we studied the impact of large-scale MoD system deployment on urban traffic patterns using multi-agent simulation. 

We ran a one-day simulation of a hypothetical large-scale MoD system that serves trips that represent the transportation demanded currently realized by private vehicles in Prague. We analyzed the results and compared the road utilization in the current state (conventional system) and in the MoD system.

Simulation results clearly showed that there would be an increase in both total traveled distance and traffic density after switching from conventional vehicles to on-demand vehicles. 
The total share of empty trips (i.e., drop-off, pick-up, and rebalancing) on traveled distance was 37.3\% with rebalancing accounting for 13.6\%. The total share of these trip types on traffic density on the roads with traffic density greater than 50\% of critical density was very similar -- 38.7\%, but here the rebalancing contributes only 6.9\%.

These results confirmed the hypothesis that the rebalancing does not contribute to congestion as much as it contributes to traveled distance because rebalancing vehicle drives against passenger-carrying vehicles, but still, the impact of the rebalancing on congestion is significant. 
Moreover, the results exposed the importance of the negative impact the station base model has on congestion. 
This is probably caused by high concentration of the pick-up and drop-off trips around the stations, and thus it can be reduced by increasing the number of the stations or by using free-floating fleet.

In future, we will focus on the development of more sophisticated congestion models, validation using data from transit sensors, and investigation of the potential of ride sharing to reduce road utilization. 

\section*{Acknowledgment}
This work was supported by the Grant Agency of the Czech Technical University in Prague, grant No. SGS16/235/OHK3/3T/13.

\bibliographystyle{unsrt}
\bibliography{bib,Zotero}

\end{document}